\documentclass[12pt]{article}
\usepackage{amsfonts}
\usepackage{graphics}

\newcommand{\rf}[1]{(\ref{#1})}
\newcommand{\beq}{\begin{equation}}
\newcommand{\eeq}{\end{equation}}
\newcommand{\bea}{\begin{eqnarray}}
\newcommand{\eea}{\end{eqnarray}}

\newcommand{\e}{\mbox{e}}
\renewcommand{\d}{\mbox{d}}

\newcommand{\tr}{\mathrm{Tr}\,}
\newcommand{\ra}{\rangle}
\newcommand{\la}{\langle}
\newcommand{\prt}{\partial}

\newcommand{\equ}{\!=\!}

\newcommand{\dg}{\dagger}
\newcommand{\oh}{\frac{1}{2}}

\newcommand{\eq}{\begin{equation}}
\newcommand{\eqx}{\end{equation}}
\newcommand{\eqn}{\begin{eqnarray}}
\newcommand{\eqnx}{\end{eqnarray}}
\newcommand{\f}[2]{\frac{#1}{#2}}
\newcommand{\eps}{\varepsilon}
\newcommand{\dl}{\delta}

\newcommand{\RR}{{\mathbb R}}

\renewcommand{\AA}{{\cal A}}
\newcommand{\HH}{{\cal H}}

\newcommand{\cor}[1]{\left\langle{#1}\right\rangle}

\begin{document}

\begin{center}
\vspace{24pt}
{ \large \bf The emergence of noncommutative target space in
noncritical string theory}

\vspace{30pt}

{\sl Jan Ambj\o rn}$\,^{a,c}$ and {\sl Romuald A. Janik}$\,^{b}$,

\vspace{24pt}
{\footnotesize

$^a$~The Niels Bohr Institute, Copenhagen University\\
Blegdamsvej 17, DK-2100 Copenhagen \O , Denmark.\\
{ email: ambjorn@nbi.dk}\\

\vspace{10pt}

$^b$~Institute of Physics, Jagellonian University,\\
Reymonta 4, PL 30-059 Krakow, Poland.\\

\vspace{10pt}

$^c$~Institute for Theoretical Physics, Utrecht University, \\
Leuvenlaan 4, NL-3584 CE Utrecht, The Netherlands.\\

\vspace{10pt}
}

\vspace{48pt}

\end{center}


\begin{center}
{\bf Abstract}
\end{center}

\vspace{8pt}
\noindent
We show how a noncommutative phase space appears in a natural
way in noncritical string theory, the noncommutative deformation
parameter being the string coupling.

\newpage

\section{Introduction}

String theory should provide us with the dynamics and
geometry of target space. In fact it should allow us to
understand precisely what target space {\it is} and to
what extent the classical concepts of space and time are
valid. The discovery of branes and dualities
in string theory has made this a vast topic.

Noncritical string theories with $c\leq 1$ defined via matrix models
allow us to study various nonperturbative  aspects of string theory.
In particular they offer us a possibility to study the summation
of string perturbation theory, which was why they attracted so much
attention in the early 90ties.
They play a role  similar to the one that certain 2d quantum field theories
play in the study of 4d quantum field theories: the possibility
to solve them allows us to test nonperturbative concepts.

More recently the study of noncritical strings
has again received much attention since
they have a worldsheet formulation as conformal field theories coupled to
Liouville field and an explicit matrix model formulation. Important progress
has been made in both formulations in the understanding of
$D$-branes and has resulted in a fruitful interplay
between the two "dual" theories.
In particular for $c<1$, where a target space formulation
had been missing, the study of $D$-branes gave us an understanding of the
target space via the study of the moduli space of the D-branes.
The space of the so-called FZZT-branes is labeled by a single
parameter $x$ and the disk amplitude $S(x)$ satisfies the
following (semiclassical) equation \cite{ss1,ss2}
\eq
\label{e.rs}
T_l(W/C)=T_k(x),~~~~~W=\partial_x S(x)
\eqx
in the $(x,W)$ plane, where $C$ is a constant and $T_p$ are Chebyshev
polynomials. 
Eq.\ (\ref{e.rs}) is valid for a minimal $(l,k)$ string
theory in the so-called conformal closed string background.
For future reference we note that in the special case of
a minimal $(2,2m-1)$ string theory, eq.\ \rf{e.rs} can be written as
\beq\label{e.rs_b}
W^2 = \Big(P_m(x,u)\Big)^2 (x+u),
\eeq
where $P(x,u)$ is a polynomial of degree $m-1$ in $x$ and $u$.
$u= \sqrt{\mu}$ where $\mu$ can be viewed as an effective
cosmological constant and $x$ can be interpreted as a boundary cosmological
constant. We also note that \rf{e.rs_b} is valid for
any so-called closed string background, $u$ being a function
$u(t)$ of a certain number of coupling constants $t_i$, $i=0,\ldots,m$,
as will be explained later.

From this point of view the semiclassical moduli space
is a rather complicated Riemann surface (\ref{e.rs_b}) and in the
ordinary complex plane $x$ the function 
$S(x) $ has cuts. This picture is true to
all finite orders in string perturbation theory. In fact it is essential
for the whole philosophy of string perturbation theory in the
context of matrix models that the cut structure in the complex plane
is unchanged. However, since it was first realized that the string
perturbation theory for non-critical strings could be resummed
it was also realized that the resummed solution to string theory
(the non-perturbative partition function in the presence of the FZZT
brane labeled by $x$) was an entire function
in the complex $x$-plane \cite{david}.
The cut had disappeared and it was seemingly
an artifact of the perturbative expansion. This point was reemphasized
in a recent paper \cite{mmss} in the context of eq.\ (\ref{e.rs}).

The purpose of this note is to explore what kind of geometrical
structure should replace the Riemann surface (\ref{e.rs_b}) on the
(nonperturbative) quantum level. In particular one has to give a
meaning to the $W$ variable on the nonperturbative level since it does
not make sense to consider {\em disk} amplitudes then. 
Promoting $W$
to an independent coordinate is also natural from the point of view of
the link between $c<1$ strings and B-model topological string on the
Calabi-Yau manifold
\eq
zw=W^2-P^2(x) (x+u)
\eqx  
where $W$ and $x$ (and $z$, $w$) are (complex) coordinates. In this
paper, however, we will not address this issue and stay purely within
the context of $c<1$ strings. 
The `quantum' $W$ and $x$ coordinates are in a completely natural way
noncommutative while the vestige of the Riemann surface (\ref{e.rs_b})
on the quantum level is an additional projection operator. 

In this note we make the above explicit -- we point out that the disappearance
of the Riemann surface \rf{e.rs} has a natural interpretation
in terms of noncommutative geometry: For a non-zero
string coupling one can define the concept of
noncommutative phase space and in some sense, to be
specified below, the Riemann surface
\rf{e.rs} appears in the limit of vanishing string
coupling which is also the limit where the noncommutative
parameter goes to zero. The formulation also has the advantage that
it brings the variables $x$ and $W$ in \rf{e.rs} on an equal footing
as phase space variables, it describes in a natural
way the "quantum mechanical" nature of target space and
it connects to a similar
(well known) description in the case of the $c=1$ string \cite{wadia}.

The integrable KdV structure of non-critical string theories
is related to a formulation in terms of free fermions (as is the
case for most integrable systems) and it is this fermionic structure
which allow us in a natural way to introduce a noncommutative phase space.

In the next section we will present the noncommutative geometric structure,
and we then study the explicit realization in $(2,2m-1)$
non-critical string theory.

\section{The noncommutative geometry framework}\label{s.ncg}

The $(2,2m-1)$ models  have a representation via
one-matrix integrals. Before the continuum limit is obtained
we can write the partition function as
\eq\label{ja0}
Z= \int dM \; \e^{-\f{1}{g_s} \tr V(M)}
\eqx
for a suitable potential $V(M)$. This matrix model can be solved using
orthogonal polynomials $P_n(x)$ satisfying
\eq
\int dx \;P_n(x) P_m(x)\; \e^{-\f{1}{g_s} V(x)} =\dl_{nm}
\eqx
A free fermion description can be obtained by introducing
fermions with wave functions
\eq
\label{e.wave}
\psi_n(x)=P_n(x) \; \e^{-\f{1}{2g_s} V(x)},
\eqx
i.e.\ we introduce a second quantized free fermion field
\beq\label{ja1}
\Psi (x) = \sum_{n=1}^\infty a_n \psi_n(x), ~~~~~~\{a_m,a_n^\dg\}=\delta_{m,n}.
\eeq
The Fermi sea is defined by filling up the first $N$ levels, where
$N$ is the size of the matrix, i.e.\
\beq\label{ja2}
a_n|0\ra = 0 ~~~{\rm for}~~n \geq N,~~~~~~~a^\dg_n|0\ra = 0~~~{\rm for}~~n <N.
\eeq
Expectation values of traces of the matrix $M$ can now be calculated nicely
in the free fermionic theory, i.e.\
\beq\label{ja3}
\la \tr M^k \ra = \la N| \Psi^\dg \hat{x}^n \Psi |N\ra,
\eeq
where $\hat{x}$ is the multiplication operator corresponding to $x$.
The origin of the fermionic nature of $\Psi$ is the Vandermonde determinant
resulting from the integration over the angular part of the matrix variables
in the integral \rf{ja0}.

The wave function of the first unoccupied level is $\psi_N(x)$, which
by the Heine formula for orthogonal polynomials can be written:
\eq\label{ja4}
\psi_N(x)= \cor{\det(x-M)},
\eqx
where the expectation value is taken with respect to partition function $Z$
in \rf{ja0}. The continuum limit of the $(2,2m-1)$ matrix model
is obtained by taking a certain
scaling limit when $N \to \infty$ and the variable $x$ is close
to the cut of eigenvalues of the matrix model in the large $N$ limit.
More explicitly one writes:
\beq\label{ja4a}
x\to x_c + a x, ~~~\f{n}{N} = 1- a^m(t_0-\tau),~~~
N\,a^{m+\oh}= \f{1}{g_s}.
\eeq
With this notation $\tau \to t_0$ means $n \to N$, and as will be described
below, $t_0$ has the
interpretation as the coupling constant related to
a perturbation with the most dominant primary operator
of the $(2,2m-1)$ conformal field theory coupled to gravity, while
$x$ has the interpretation of a boundary cosmological constant.

In this limit one obtains (see for instance \cite{moore} for a review)
\beq\label{ja5}
\psi(x_c+ a x) \propto \psi(\tau,x),
\eeq
where the double-scaled function $\psi(\tau,x)$ satisfies
\beq\label{ja6}
Q \, \psi(\tau,x)=x \psi(\tau,x),~~~
Q= {g_s^2} \f{\d^2}{\d \tau^2} -u(\tau,g_s).
\eeq
Eq.\ \rf{ja6} can be viewed as a Schr\"{o}dinger-like equation
with $-Q$ playing the role of the Hamiltonian, $u(\tau)$ the role
of the potential and $-x$ the role of energy. It would  allow for a
standard WKB expansion in $g_s$ if it was not for the fact that the
"potential" $u(\tau,g_s)$ was itself a function of $g_s$, determined
from the so-called string equation
\beq\label{ja7}
[Q,P]=g_s,~~~~~P\psi(\tau,x) = g_s\f{\d}{\d x} \psi(\tau,x)
\eeq
where $P$ is a polynomial in $d/d\tau$ and $u$.
As is seen $Q$ and $P$,
represent in a  way the noncommutative phase space variables $q=x$ and
$p= g_s d/dx$. $p$ is then to be identified with the $W$ coordinate on
the (nonperturbative) quantum level.
However, from \rf{ja6} another (conceptually simpler)
representation of a noncommutative phase space will emerge which we now
turn to describe.

The important point for us at this stage is the notion of a Fermi sea,
the ground state, in a theory of free fermions, above which we consider
excitations. We thus have to project from wave functions all components
below the Fermi level, which is explicitly done by the kernel
\eq\label{ja8}
K_{c<1}(x,y)=\sum_{n=N}^\infty \psi_n(x)\psi_n(y)
\eqx
It can be written as
\beq\label{ja8a}
K_{c<1}(x,y) =\dl(x-y)-
\sum_{n=0}^{N-1} \psi_n(x)\psi_n(y)= \dl(x-y)-K_{rmm}(x,y)
\eqx
where
\eq\label{ja9}
K_{rmm} \propto \f{\psi_{N}(x) \psi_{N-1}(y)-\psi_{N-1}(x)\psi_{N}(y)}{x-y}
\eqx
is the standard random matrix kernel where an appropriate
double scaling limit should be taken according to \rf{ja4a}.
Clearly a derivative with respect to $\tau$ is involved
in the double scaling limit of \rf{ja9} and using eq.\ \rf{ja6}
one obtains
\beq\label{ja9a}
K_{c<1}(x,y;t_0)  = \int_{t_0}^{\infty} \d \tau \; \psi^*(\tau,x)\psi(\tau,y)
\eeq
The first unoccupied level is then exactly the
Baker-Akhiezer function (or the 1-point function of the FZZT brane)
as is clear from \rf{ja4} and in the double scaling
limit we can write $\psi(t_0,x)$.

The above construction fits nicely into the general framework
of non-commutative geometry. Let the starting point
be the noncommutative plane.
It can be defined as the algebra $\AA$ generated by $p$ and $q$
operators acting on $\HH=L^2(\RR)$.

This Hilbert space in the context of $c<1$ strings is naturally
identified with the space of wavefunctions (\ref{ja5}) of the
excitations of the random matrix model in the double scaling limit (we
will discuss the Fermi sea in this context shortly).

In order to pass to the picture of functions of two
variables (these will be the $p\equiv iW$ and $q\equiv x$ of the target
space of the minimal string theory) with noncommutative multiplication
one uses the Weyl map 
\eq\label{ja11}
f(p,q) \longrightarrow \hat{f}= \int\f{\d k_1\d k_2 }{(2\pi)^2}\;
\e^{i (k_1 \hat{q}+k_2 \hat{p})} \; \tilde{f}(k_1,k_2),
\eqx
where $\tilde{f}$ is the Fourier transform of $f$.
Eq.\ \rf{ja11} maps functions to operators, and defines the noncommutative
multiplication by
\eq\label{ja12}
\hat f \circ \hat g \to \longrightarrow f * g.
\eqx
The inverse map, sometimes called the Wigner map, is constructed
by introducing via \rf{ja11} the operator $\hat{\Delta}$
corresponding to the delta-function
$\delta(q-q_0)\delta(p-p_0)$:
\beq\label{ja13}
\hat{\Delta}(q_0,p_0)=
\int\f{\d k_1\d k_2 }{(2\pi)^2}\;
\e^{i (k_1 \hat{q}+k_2 \hat{p})} \; \e^{-i(q_0k_1+p_0k_2)},
\eeq
and it is
\beq\label{ja14}
\hat{f} \longrightarrow f(q,p)= \tr \Big(\hat{f} \;\hat{\Delta}(q,p)\Big)
\eeq
The noncommutative product defined in this way by \rf{ja12}
is exactly equivalent
to the standard Moyal product:
\eq\label{ja15}
f(q,p) * g(q,p) = f(q,p) \,\exp\left(i\,\frac{1}{2} \,
\overleftarrow{\partial_\mu} \theta_{\mu\nu}
\overrightarrow{\partial_\nu}\right)\,g(q,p)
\eqx
where $\mu,\nu$ refers to the $q,p$  coordinates and $\theta_{\mu\nu} =
\theta \eps_{\mu\nu}$ is the noncommutative parameter of the the theory.

At this stage we see that one ingredient of the $c<1$ string theory
has not yet been incorporated so far. Since the elementary excitations
are fermions and all the levels in the Fermi sea are already filled,
the excitations $\psi(\tau,x)$ cannot probe the whole Hilbert space
$\HH=L^2(\RR)$ and one has to project out from $\HH$ the states in the
Fermi sea through (\ref{ja8}). This projection leads to an additional
geometrical structure in the noncommutative plane which is essentially
the quantum nonperturbative analog of the classical Riemann surface
(\ref{e.rs_b}) as we will show.

To this end let us note that associated with a given quantum wave
function $\psi(x)$ we have the projection operator $|\psi \ra \la
\psi|$ and the Wigner map applied to this operator gives the Wigner
function $f_\psi(q,p)$ associated with the wave function $\psi(x)$:
\beq\label{ja16}
f_\psi(q,p) = \int \f{\d x}{2\pi \hbar} \; \e^{ip x/\hbar} \;
\psi^*(q-x/2)\psi(q+x/2).
\eeq
where $q=(x_1+x_2)/2$, the variables $x_1$ and $x_1$ being the
coordinates which appear in the projector $\la x_2|\psi\ra\la \psi|x_1\ra$.
Now we are ready to write the Wigner map associated with the projector
\beq\label{ja17}
\hat{\rho} = \sum_{n=N}^\infty |\psi_n\ra\la \psi_n|
\eeq
which leads to the associated Wigner function
\beq\label{ja17a}
f_\rho(q,p)= \int \f{\d x}{2\pi \hbar} \; \e^{ipx/\hbar} \;
\rho(q-x/2,q+x/2),
\eeq
where again $\rho(x_1,x_2) = \la x_1| \hat{\rho}|x_2\ra$ and $q=(x_1+x_2)/2 $.

The projection acts then on functions of two variables $q=x$ and $p=iW$
by the star multiplication:
\eq
f(q,p) \longrightarrow f_\rho(q,p) * f(q,p)
\eqx 

The geometrical structure which appears in the
examples considered in this paper
supplements the noncommutative plane by an appropriate (orthogonal)
projection operator $\hat{K}$.
One then truncates the space of functions to the image of $\hat{K}$:
\eq
\HH_K=\{ \hat{K}f : f\in L^2(\RR) \}
\eqx
and similarly the algebra of operators acting on $\HH_K$:
\eq
\AA_K \equiv \{\hat{K}\hat{O} : \hat{O} \in \AA \} =
\{\hat{K}\hat{O}\hat{K} : \hat{K} \in \AA \}
\eqx

It is seen that this structure is precisely the one present
also in $c=1$ non-critical string theory \cite{wadia,sen,aj}.
Let us now discuss in more detail the implementation
of this abstract noncommutative geometrical
structure in the $(2,2m-1)$ minimal models.

Before we proceed with the general discussion let us illustrate the
above notions in the simplest example of $(2,1)$ strings.

\subsubsection*{$(2,1)$ minimal string theory}

In this case the Riemann surface is
\eq
\label{e.curve}
W^2-q-t_0=0
\eqx
Let us now proceed to find the projector which represents the geometry
of the $(2,1)$ theory according to (\ref{ja17a}). In this case the
double scaling wavefunctions can be obtained exactly and we have
\beq\label{ja39}
\Big(\hbar^2\f{d^2}{dt_0^2} -t_0\Big) \psi(t_0,x)= x \psi(t_0,x),~~~~~
\psi(t_0,x) =  {\rm Ai}\Big( \f{x+t_0}{\hbar^{2/3}}\Big),
\eeq
while the Wigner transform is
\beq\label{ja41}
f_\psi(q,p) = \frac{1}{\pi \hbar^{2/3}} \;
{\rm Ai}\Big(\f{p^2+(q+t_0)}{\hbar^{2/3}}\Big)
\eeq
We see already here that the curve (\ref{e.curve}) appears as the
argument of the Airy function.
The Wigner transform of the projector is then obtained by integrating
over $t_0$
\beq\label{ja42}
f_\rho(q,p) = \int^{\infty}_{t_0} \d t'_0 f_\psi(q,p;t'_0),
\eeq
Let us now examine the classical limit $\hbar \to 0$.
Using the known asymptotics of integrals of Airy functions
we have in the classically forbidden region
\beq\label{ja43}
f_\rho(q,p) \sim \f{\hbar^{1/2}}{(p^2+q+t_0)^{3/4}} \; \exp\Big(-\f{2}{3\hbar^{2/3}}
(p^2+q+t_0)^{3/2}\Big),
\eeq
while in the classically allowed region
\beq\label{ja44}
f_\rho (q,p)\sim 1-\f{\hbar^{1/2}}{\sqrt{\pi}(-p^2-q-t_0)^{3/4}}
\cos\Big(\f{2}{3\hbar^{2/3}}(-p^2-q-t_0)^{3/2}+\f{\pi}{4}\Big).
\eeq
Therefore the projector in this limit approaches the Heaviside step
function
\eq
f_\rho(q,p) \longrightarrow \theta(p^2+q+t_0)
\eqx
The Riemann surface curve appears then as the boundary of the
classically allowed region which is natural since the FZZT 1-point
function is the lowest excitation above the Fermi sea.

We will now proceed to explore these issues in the much more
complicated case of $(2,2m-1)$ minimal string theories. 

First we outline how one  obtains
\rf{e.rs} in the semiclassical limit where $\hbar \equiv g_s \to 0$
from the Wigner-function corresponding to $\psi(t_0,x)$,
{\it viewed as a function of $x$, not $t_0$}.

\section{The semiclassical limit and beyond}

The semiclassical solution to $(2,2m-1)$ non-critical strings is
entirely determined by the disk-amplitude. It can be written
as function of a number of coupling constants $t_i$, $i=0,m$,
and the boundary cosmological constant $q$. For a given choice of
$t_i$ the explicit formula for  the
semiclassical disk-amplitude is
\cite{book}:
\beq\label{ja18a}
S(q,t) = \oh\oint_\infty \frac{\d \Omega}{2\pi i}\;V'(t,\Omega)
\;\log
\left(
\f{1+\sqrt{\f{q+u(t)}{\Omega+u(t)}}}{1-\sqrt{\f{q+u(t)}{\Omega+u(t)}}}
\right).
\eeq
while the formula for the derivative of the disk amplitude
w.r.t. the boundary cosmological constant is (using the string equation
\rf{ja20})
\beq\label{ja18}
W(q,t)=\f{\prt S}{\prt q}=\oh \left(  \oint_\infty \frac{\d \Omega}{2\pi i}\;
\f{V'(t,\Omega)}{(\Omega-q)\sqrt{\Omega+u(t)}} \right) \;\sqrt{q+u(t)},
\eeq
The (derivative of the) potential is defined by
\beq\label{ja19}
V'(t,q)= \sum_{k=0}^m t_k \, q^{k-1/2},
\eeq
and the  function $u(t)$ is determined by the string equation \rf{ja7}. It can
be written as a differential equation in $u$:
\beq\label{ja19a}
\sum_{k=0}^m t_k R_k(u) = 0,
\eeq
where $R_k(u)$ denotes the so-called Gelfand-Dikii differential polynomials
of $u$, considered as a function of $t_0$\footnote{The Gelfand-Dikii
differential polynomials can be recursively defined by
$$
\f{\d}{\d {t_0}} R_{k+1}(u)= \left(\f{\hbar^2}{4}\f{\d^3}{\d t_0^3}
-u\f{\d}{\d {t_0}} -\oh\f{\d u}{\d {t_0}} \right) R_k(u),~~~
R_0=1,~~R_1=-\f{u}{2}.
$$}.
In the semiclassical limit \rf{ja19a} reduces  to
\beq\label{ja20}
\sum_{k=0}^m c_k t_k  u^{k}  =0~~~{\rm or}~~~
\oint_{\infty} \f{\d \Omega}{2\pi} \; \f{V'(\Omega,t)}{\sqrt{\Omega+u(t)}}=0,
\eeq
where the $c_k$'s are the power series coefficients of $(1+x)^{-1/2}$.
In this semiclassical approximation $u(t) = \sqrt{\mu(t)}$ where
$\mu$ is the effective cosmological constant of 2d Euclidean quantum
gravity coupled to a $(2,2m-1)$ conformal field theory, expressed
via the string equation as a function of the coupling constants $t_i$.

A choice of coupling constants $t_i$ is called a choice of closed
string background and from \rf{ja18}
it is seen that the semiclassical disk amplitude can be written as
\beq\label{ja21}
W(x,t)= \oh \left( \sum_{l=0}^m t_l P_l(x,u)\right)\sqrt{x+u},
\eeq
where the polynomial multiplying $t_l$ is
\beq\label{ja22}
P_l(x,u)= \sum_{i=0}^{l-1} c_ku^k x^{l-1-k}.
\eeq
Similar remarks applies to $S(x,t)$ which has a similar
representation as \rf{ja21} with  polynomials $\tilde{P}_l(x,u)$
one degree higher than $P_l(x,u)$.

Differentiating $S(x,t)$ w.r.t. $-t_0$ reproduces precisely
the  FZZT amplitude for an insertion of the primary operator
of the most negative
(gravitationally dressed) scaling dimension \footnote{Differentiating
$S$ or $W$ after the other coupling constants $t_i$  produce
expressions, also only depending on $u$ and $q$, with no explicit
reference to $t_i$. However, in general the $t_i$'s cannot for $0<i\leq m-2$
be identified directly with coupling constants related to
the primary operators of scaling dimension $-(m-2+i)/2$ due to
operator mixing with the (gravitational) descendents of the
operator corresponding to $t_0$.},
$\Delta_{r\equ 1,s\equ m-1}=-(m-2)/2$:
\beq\label{ja23}
\f{\d S(x,t)}{\d t_0} = {\sqrt{x+u}}.
\eeq
We note that this formula implies that
\bea\label{ja24}
S(x,t) &=& \int_{t_0(x)}^{t_0} \d t_0'\;
{\sqrt{x+u(t'_0,t_1,\ldots)}} \\
W(x,t) &=& \oh \int_{t_0(x)}^{t_0} \d t_0'\;
\f{1}{\sqrt{x+u(t'_0,t_1,\ldots)}},\label{ja24a}
\eea
where $u(t_0(x)) = -x$.

A full non-perturbative treatment involves the following steps:
first solve the string equation \rf{ja19a} for $u$. Then solve
the Schr\"{o}dinger equation \rf{ja6} for $\psi(\tau,x)$ and
finally construct the projector \rf{ja9a}. The Wigner transform
either of $\psi(\tau,x)$ or the projector $K(x,y;\tau)$
then provides us with variables $(q,p)$ which can be viewed as
phase space variables which lives on a non-commutative phase space
defined by the projector as described above. Let us now show that
in a naive semiclassical ($\hbar\equiv g_s \to 0$) limit we obtain
formally the classical phase space $p=W(q)$, i.e.\ for the
the choice of $t_k$'s corresponding to a conformal background precisely
\rf{e.rs}.

Let us write the solution to \rf{ja6} as:
\beq\label{ja25}
\psi(t_0,x)=\sqrt{R(x,u)}\; \e^{\pm S(x,t_0)},~~~~
\f{\prt S}{\prt t_0}=\frac{1}{R}.
\eeq
The \rf{ja6} can be written as
\beq\label{ja26}
\f{\hbar^2}{4}\left( \left(\f{\d R}{\d t_0}\right)^2-
2 R \f{\d^2R}{\d t_0^2}\right)
+(x+u(t_0)) R^2 =1.
\eeq
Eqs.\ \rf{ja25} and \rf{ja26} is one version of
the WKB-equations\footnote{Note that
if we differentiate \rf{ja26} with respect to $t_0$ we obtain
$$
\left(\f{\hbar^2}{4} \f{\d ^3}{\d t_0^3} -(x+u)\f{\d }{\d t_0} -
\oh\f{\d u}{\d t_0}\right) R=0,
$$
which shows that $R(x,u)$ is the generating functional for the Gelfand-Dikii
polynomials:
$$
R(x,u) = \sum_{k=0}^\infty R_k(u) x^{-k-1/2}.
$$
It also follows from the definition that $R(x,u)$ is the diagonal element of
the resolvent: $\la t_0|(Q-x)^{-1}|t_0\ra$.},
which are equivalent to the Schr\"{o}dinger equation \rf{ja6}.
In the WKB {\it approximation} where only the lowest order
terms in $\hbar$ is kept in $R$ and $S$  we have from \rf{ja26}
\beq\label{ja27}
R(x,u)= \f{1}{\sqrt{x+u}},
\eeq
and we obtain $S$ from \rf{ja25} by integrating from the classical
turning point, i.e.\ the point $t_0(x)$ where $u(t_0(x),t_1,\ldots)=-x$,
\beq\label{ja28}
S(x,t)=\int_{t_0(x)}^{t_0} \d t_0' \;\sqrt{x+u(t_0',t_1,\ldots)}.
\eeq
Comparing with \rf{ja24} we see explicitly that the WKB approximation indeed
reproduces the exponential of the semiclassical disk amplitude,
i.e.\ the partition function of a single FZZT brane, as
expected from the general discussion above.

\subsection{The naive limit}

We can now use this WKB approximation to calculate
the Wigner transform:
\beq\label{ja29}
f_\psi(q,p)= \int \f{\d x}{2\pi \hbar} \; \e^{i p x/\hbar}
\psi^*(q-x/2)\psi(q+ x/2).
\eeq
We emphasize that this Wigner transformed function is different
from the ordinary Wigner function corresponding to $\psi_x(t_0)$ considered
as an eigenfunction to the Schr\"{o}dinger operator $Q$ with
eigenvalue $x$. We perform the transformation in $x$, not in $t_0$,
which is the variable appearing in $Q$.

Since $\hbar \to 0$ it is {\it tempting} (but not entirely correct as
will be discussed shortly) to change integration variable $x= \hbar y$ and use the
following approximations in the integrand:
\beq\label{ja30}
S(q+\hbar y/2) = S(q)+\f{\hbar y}{2} \, \f{\prt S(q)}{\prt q}+ O(\hbar^2),
\eeq
and $R(q+\hbar y/2,u)R^*(q-\hbar y/2) =|R(q,u)|^2+O(\hbar^2) $.
If we first consider $q$ which is in the  "classically allowed region" 
(where $S(q)$ is imaginary), and use eq.\ \rf{ja30} in eq.\ \rf{ja29}
we obtain
\beq\label{ja31}
f_\psi(q,p) = | R(q,u(t))| \;
\dl\Big(p^2+\Big(\f{\prt S(q,t)}{\prt q} \Big)^2\Big)
= \dl(t_0-t_0^{i}(q,p)),
\eeq
where $t_0^{i}(q,p)$ is defined as the value of $t_0$ where the
argument of the $\dl$-function is zero for given, fixed
values of $p$ and $q$. In the ``classical forbidden region''
(where $S(q)$ is real) we obtain
in the same approximation zero because the contribution is exponentially
suppressed by $\exp(-2S(q)/\hbar)$ and $\hbar \to 0$.

From eq.\ \rf{ja31} we finally obtain for the Wigner-transformed
projector (see \rf{ja17a} and \rf{ja9a})
\beq\label{ja32}
f_\rho (q,p) = \theta\Big(p^2+\Big(\f{\prt S(q,t)}{\prt q}\Big)^2 \Big).
\eeq

In the above semiclassical approximation  we have thus seen that
the Wigner transform gives us the classical phase-space
\rf{e.rs_b}, provided we make the identification $W \to ip$.
This identification is indeed natural since the $W$ appearing
in \rf{e.rs_b} should be viewed as  a classical representative
of the $P$ appearing in \rf{ja7}, i.e.\ a representation
of the operator $d/dx$. Clearly, by construction, the $p$ appearing
in the Wigner transform is a ``real'' $p$, related to $-i d/dx$.

\subsection{The WKB Wigner function}

\begin{figure}[t]
\centerline{\scalebox{0.6}{\rotatebox{0}{\includegraphics{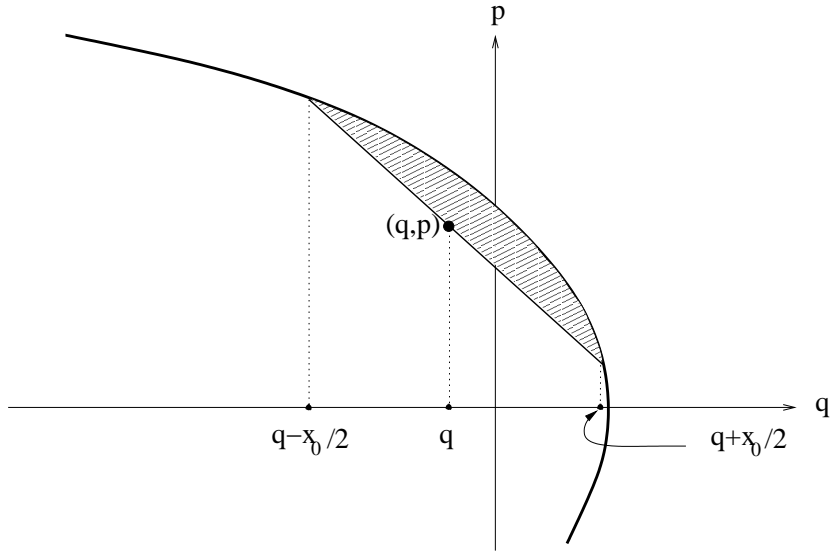}}}}
\caption[fig1]{The graphical representation of the 
solution to eq.\ \rf{ja33}, the shaded area $A(q,p)$ 
being defined in eq.\ \rf{ja34a}.} 
 \label{fig1}
\end{figure}

The approximation \rf{ja30} does not commute unfortunately with the
integration in \rf{ja29}. Thus the results \rf{ja31} and \rf{ja32}
are not correct, although they indeed lead to the correct semiclassical
expectation values of nice observables like $\hat{x}^n$. A saddle-point
calculation of \rf{ja29} is however quite reliable except when the
values of $q$ and $p$ are very close to the classical phase space
orbit defined by \rf{ja32}. From \rf{ja29} we obtain the saddle point
equation for $x$:
\beq\label{ja33}
W(q+x_0/2)-W^*(q-x_0/2)=2ip
\eeq
In the classically allowed region (where $q+u(t) < 0$) $W(q)$ is imaginary
and eq.\ \rf{ja33} allows for ordinary quantum mechanical
systems a standard solution (see \cite{heller,berry,habib}) since the
phase space curve is concave (see fig.\ \ref{fig1}), as emphasized in
\cite{heller,berry,habib}. 
The saddle-point solution is
\beq\label{ja34}
f_\psi(q,p)= c^2 \, \left(F(q,p)\;
 \exp \Big(\f{i}{\hbar} A(q,p)\Big) +
{\rm c.c.}\right)
\eeq
where $c^2$ is a normalization factor irrelevant for the
present discussion.  The prefactor $F(q,p)$ is
\beq\label{ja34b}
F(q,p)=\f{\sqrt{R(q+\f{x_0}{2},t)
R^*(q-\f{x_0}{2},t)}}{\sqrt{W'_{x_0}(q+\f{x_0}{2})+
{W'_{x_0}}^*(q-\f{x_0}{2})}},
\eeq
where $x_0$ is the solution to \rf{ja33} and
$W'_{x_0}$ denotes the derivative after $x_0$. Finally
$A(q,p)$ is the value of the exponent in \rf{ja31} for the
solution $x_0$ of \rf{ja33}, i.e.\
\beq\label{ja34a}
A(q,p) =px_0+S(q+\f{x_0}{2})+S(q-\f{x_0}{2}).
\eeq
For a concave phase space orbit $A$ and $(q,p)$ in the
"interior", $A$ clearly has the interpretation
of the area shown in fig.\ \ref{fig1}.
If $W$ is imaginary then
$-x_0$ is also a solution. Eq.\ \rf{ja34} includes both
saddle points via the complex conjugate term and
$f_\psi(q,p)$ is real and oscillating. {\it Note that these oscillations
do not disappear in the limit $\hbar \to 0$.} For an ordinary
quantum mechanical system with $(q,p)$ outside the classically allowed
region in phase space corresponding to a certain energy eq.\ \rf{ja33}
will have a complex solution $x_0$ and $f_\psi(q,p)$ will fall off exponentially.

If eq.\ \rf{ja33} holds for values of $(q,p)$ where $x_0\to  0$, i.e.\ when
we approach the "classical phase space" defined by \rf{ja32}, we see
that \rf{ja34} diverges due to the divergence of the
prefactor $F(q,p)$ for $x_0 \to 0$. This is not related to the question of the
reliability of the WKB approximation of $\psi(t_0,x)$ at
the classical turning point, but just to a failure of the
quadratic saddle point calculation of the Wigner integral
and one can improve the calculation by the method of
uniform approximation, which involves in a natural way the Airy functions.
This was done in \cite{berry} where
a closed expression was given, which is not only valid near the classical
turning points but in the whole phase space, and which reduces
to \rf{ja34} away from the classical orbit:
\beq\label{ja35}
f_{\psi}^{wkb}(q,p)=\f{c^2}{\hbar^{2/3}} \,F(q,p)\; A^{\f{1}{6}}(q,p)
\;{\rm Ai} \left( -\Big[\f{3A(q,p)}{2\hbar}\Big]^{2/3}\right)
\eeq
where $c$ is a normalization constant {\it independent} of $\hbar$,
irrelevant for the present discussion. The
Wigner WKB-function is a good approximation to the exact result
for small $\hbar$, in the sense that it incorporates already
the improvements to the simple WKB approximation present in
the so-called uniform semiclassical approximation even if the
starting point was the simple WKB-approximation for $\psi$ \cite{berry}.
In particular, it is finite at the classical orbit except in exceptional
cases discussed in \cite{berry}.

Again, the non-disappearance of
$f_{\psi}(q,p)$ in the classically allowed region but away from the classical
orbit is quite manifest in the expression \rf{ja35}. The following
features are noteworthy: (1) the peak of $f_\psi(q,p)$ grows as $\hbar^{-2/3}$
as $\hbar \to 0$. Thus $f_\psi(q,p)$ {\it a priori} had a chance to reproduce
the $\dl$-function at the classical orbit \rf{ja31}. However the
Airy function oscillates in this region and these oscillations also
diverge in the limit $\hbar \to 0$ although slightly weaker.
Folded with sufficiently smooth functions these oscillations
average out, and for "observables" $O(q,p)$ corresponding to
such functions the $\hbar \to 0$ limit of $f_\psi(q,p)$ will indeed be given
by \rf{ja31} (see \cite{heller} for 
discussion and examples). (2) Since the peak of the Airy function is not at
zero it follows from \rf{ja35} that the "fuzzy classical orbit"
is actually displaced a distance proportional to $\hbar^{2/3}$ and
the spread of this fuzzy orbit is also of order $\hbar^{2/3}$.

\subsection{Beyond the semiclassical limit (almost)}

The simplest example is provided by the (2,1) minimal string model ($m
\equ 1$). 
In this case we have  simply $u(t)= t_0$. Thus we can explicitly solve
the system both in the semiclassical limit and completely non-perturbatively
in $\hbar$.

Let us first state form of the WKB-approximation above. Let us denote
$-t_0$ by $E$, the "classical energy". Thus we have on a classical orbit
\beq\label{ja36}
p_{cl}^2+q_{cl} = E,~~~~ \frac{3}{2} A(q,p)=(E-(p^2+ q))^{3/2},
\eeq
while $F(q,p) \propto A^{-1/6}(q,p)$, and we obtain from \rf{ja35}
\beq\label{ja37}
f_\psi^{wkb}(q,p)\propto \f{1}{\hbar^{2/3}}\;{\rm Ai}
\Big( \f{p^2+q+t_0}{\hbar^{2/3}}\Big).
\eeq

The exact solution was already cited earlier (\ref{ja39}) and its
Wigner transform is
\beq
f_\psi(q,p) = \frac{1}{\pi \hbar^{2/3}} \;
{\rm Ai}\Big(\f{p^2+(q+t_0)}{\hbar^{2/3}}\Big)
\eeq
Maybe not surprisingly the Airy-function uniformized WKB approximation
is exact for the linear potential! However, it suggests that many of the
lessons one can draw from \rf{ja35} might well be valid beyond the
semiclassical approximation. At least here we have seen that in the
context of the simplest non-critical string theory it includes
a summation over all genera.

Given the explicit expression \rf{ja41} we can calculate the
Wigner transform $f_\rho(q,p)$ of the projector  $K$.
The projector $f_\rho(q,p)$ is better behaved than
$f_\psi(q,p)$ w.r.t. the classical limit.
As shown in (\ref{ja43})--(\ref{ja44}) the projector $f_\rho(q,p)$
indeed approaches the naive limit \rf{ja31} for $\hbar \to 0$.

This argument can clearly be extended to the general WKB Wigner transform
given by \rf{ja35}. Integrating with respect to $t_0$ will produce
additional factors of $\hbar$ which will suppress the oscillations
in the "classically allowed region" in the limit $\hbar \to 0$.

\section{Discussion}

Non-critical string theory has always been a useful
laboratory for the study new ideas in string theory.
Most recently it has received renewed attention after
it was realized that the concepts of branes and open-closed
string duality could be analyzed using matrix model technology.
In this connection a target space interpretation of the
the $c < 1$ non-critical strings was suggested, target space
being identified with a Riemann surface which was explicitly
constructed in the conformal background for a $(k,l)$ non-critical
string in the semiclassical approximation.

One of the virtues of non-critical string theory is that
it, at least to some extent, offers us a non-perturbative
definition of string theory. In particular for the $(2,2m-1)$
non-critical string theories, where $m$ is odd, this is relatively
concrete: one can define a theory which should represent
the non-perturbative sum over all genera in the sense
that one can actually solve the string equation and 
find a ``physically acceptable'' string susceptibility 
function $u(t)$ well defined on the whole real $t_0$-axis 
\cite{shenker}. One problem arises
in this connection: the elaborate structure of target space
seemingly present semiclassically disappears completely and target
space reduces to the complex plane. It seems a quite
unsatisfactory situation if we extrapolate back to critical
string theory. It would imply that we know nothing about
target space unless we are able to perform the sum over
all genera in critical string theory. It prompted us to
look for another target space interpretation of the
non-critical strings. We were here inspired by the
fact that in $c=1$ non-critical string theory
(where the notion of target space is more clear than for
non-critical string theories with $c <1$) the concept of
a non-commutative phase space appears naturally \cite{wadia}.

In this note we have shown that noncommutative phase space
indeed appears in a natural way in non-critical string theory.
The  Wigner function associated with the projector which defines
the non-commutative phase space reduces in the semiclassical
limit precisely to \rf{ja32}. It also makes it possible
to avoid taking about Riemann surfaces altogether, if wanted.
Target space becomes associated with a (real) coordinate $q$
to which we can associate a noncommutative "momentum" $p$.
Since we consider $c < 1$ we have no time evolution in this
target space but orbits of constant "energy" makes sense
and some aspects of the noncommutative nature
of the phase space survive even semiclassically since
the WKB approximation precisely defines the semiclassical
FZZT brane.

Let us perhaps contrast the appearance of noncommutative geometry in
these theories with the earlier studies of noncommutativity in the
context of backgrounds with nonzero $B_{NS-NS}$ and the resulting open
string theories \cite{SeibergWitten}. There the
noncommutativity was a property of a particular background and could
be turned off at will, while here it is intrinsically linked to the
string coupling constant.

Let us note that the same noncommutative geometrical structure appears
naturally also in other contexts in string theory. $c=1$ strings also
have a free fermion formulation with a filled Fermi sea whose
classical boundary is the curve $p^2-q^2-\mu=0$. In an analogous
fashion the quantum description will again be the noncommutative plane
with a projection operator which represents  a fuzzy noncommutative
version of the Fermi sea.

Since the $c<1$ and $c=1$ string theories are closely associated with
B-model topological strings on Calabi-Yau's of the form $zw-H(p,q)=0$,
the above results also suggest that similar noncommutative structures
should also appear in that context.

It would be also interesting to explore the issues of noncommutativity
in the context of the $\f{1}{2}$-BPS sector in the AdS/CFT
correspondence which on the classical level is mapped to a free
fermion description \cite{Berenstein} and the corresponding classical
geometry is encoded by the geometry of the Fermi sea \cite{LLM}. 

Finally on a much more speculative level let us mention A-model
topological strings formulated through the sum over 3D partitions
\cite{Vafa3d}. The blown-up Calabi-Yau's (associated to toric diagrams
$\sim$ 3D partitions) are represented by a lattice of points in an
octant. The missing points are very naturally identified with harmonic
oscillator eigenfunctions which are projected out\footnote{E.g. such
an interpretation for the toric diagram of $\mathbb{P}^1$ yields the
fuzzy sphere.}. The melting crystal partition function would then have
an interpretation as a summation over (admissible) quantum geometries
represented by projectors.

It would be fascinating to understand more deeply the interrelations
between all these pictures.

\subsubsection*{Acknowledgment}

RJ was supported in part by KBN grants 2P03B08225 (2003-2006) and
1P03B02427 (2004-2007).

\bigskip

\noindent{\bf Note added:} As this paper was being finished the work
\cite{gmr} appeared which also addresses the quantum phase space of
minimal strings.


\begin{thebibliography}{99}
                                                                             
\bibitem{ss1}
  N.~Seiberg and D.~Shih,
  {\it Branes, rings and matrix models in minimal (super)string theory},
  JHEP {\bf 0402} (2004) 021
  [arXiv:hep-th/0312170].

\bibitem{ss2}
D.~Kutasov, K.~Okuyama, J.~w.~Park, 
N.~Seiberg and D.~Shih,
{\it Annulus amplitudes and ZZ branes in minimal string theory},
  JHEP {\bf 0408} (2004) 026
  [arXiv:hep-th/0406030].

\bibitem{david}
  F.~David,
  {\it Loop Equations And Nonperturbative Effects In Two-Dimensional Quantum
  Gravity},
  Mod.\ Phys.\ Lett.\ A {\bf 5} (1990) 1019;\\
  F.~David,
  {\it Phases Of The Large N Matrix Model And Nonperturbative Effects In 2-D
  Gravity},
  Nucl.\ Phys.\ B {\bf 348} (1991) 507.

\bibitem{mmss}
  J.~Maldacena, G.~W.~Moore, N.~Seiberg and D.~Shih,
  {\it Exact vs. semiclassical target space of the minimal string},
  JHEP {\bf 0410} (2004) 020
  [arXiv:hep-th/0408039].

\bibitem{wadia}
  A.~Dhar, G.~Mandal and S.~R.~Wadia,
  {\it ``Nonrelativistic fermions, coadjoint orbits of W(infinity) and
  string field theory at c = 1,''}
  Mod.\ Phys.\ Lett.\ A {\bf 7} (1992) 3129
  [arXiv:hep-th/9207011];\\
  G.~Mandal and S.~R.~Wadia,
  {\it Rolling tachyon solution of two-dimensional string theory},
  JHEP {\bf 0405} (2004) 038
  [arXiv:hep-th/0312192].

\bibitem{moore}
 P.~H.~Ginsparg and G.~W.~Moore,
  {\it Lectures on 2-D gravity and 2-D string theory},
  [arXiv:hep-th/9304011].

\bibitem{sen}
  A.~Sen,
  {\it Open-closed duality: Lessons from matrix model},
  Mod.\ Phys.\ Lett.\ A {\bf 19} (2004) 841
  [arXiv:hep-th/0308068].

\bibitem{aj}
  J.~Ambjorn and R.~A.~Janik,
 {\it The decay of quantum D-branes}
  Phys.\ Lett.\ B {\bf 584} (2004) 155
  [arXiv:hep-th/0312163].

\bibitem{book}
  J.~Ambjorn, B.~Durhuus and T.~Jonsson,\\
  {\it Quantum geometry. A statistical field theory approach},
  Cambridge Monogr.\ Math.\ Phys.\  {\bf 1} (1997) 1.

\bibitem{heller}
E.~J.~Heller,
{\it Phase space interpretation of semiclassical theory},
Journ.\ Chem.\ Phys.\ {\bf 67} (1977) 3339.

\bibitem{berry}
M.\ V.\ Berry, 
{\it semiclassical mechanics in phase space: a study 
of Wigner's function},
 Phil.\ Trans.\ R.\ Soc. London, {\bf 287} (1977) 237.

\bibitem{habib}
  S.~Habib,
  {\it The Classical Limit In Quantum Cosmology. 1 
Quantum Mechanics And The Wigner Function},
  Phys.\ Rev.\ D {\bf 42}, (1990) 2566;\\
  S.~Habib and R.~Laflamme,
  {\it Wigner Function And Decoherence In Quantum Cosmology},
  Phys.\ Rev.\ D {\bf 42}, 4056 (1990).


\bibitem{shenker}
  M.~R.~Douglas, N.~Seiberg and S.~H.~Shenker,
  {\it Flow And Instability In Quantum Gravity},
  Phys.\ Lett.\ B {\bf 244}, 381 (1990).

\bibitem{SeibergWitten}
 N.~Seiberg and E.~Witten,
  {\it ``String theory and noncommutative geometry,''}
  JHEP {\bf 9909} (1999) 032
  [arXiv:hep-th/9908142].

\bibitem{Berenstein}
  D.~Berenstein,
  {\it ``A toy model for the AdS/CFT correspondence,''}
  JHEP {\bf 0407}, 018 (2004)
  [arXiv:hep-th/0403110].

\bibitem{LLM}
  H.~Lin, O.~Lunin and J.~Maldacena,
  {\it ``Bubbling AdS space and 1/2 BPS geometries,''}
  JHEP {\bf 0410} (2004) 025
  [arXiv:hep-th/0409174].

\bibitem{Vafa3d}
  A.~Okounkov, N.~Reshetikhin and C.~Vafa,
  {\it ``Quantum Calabi-Yau and classical crystals,''}
  [arXiv:hep-th/0309208];\\
  A.~Iqbal, N.~Nekrasov, A.~Okounkov and C.~Vafa,
  {\it ``Quantum foam and topological strings,''}
  [arXiv:hep-th/0312022].


\bibitem{gmr}
C.\ G.\ Gomez, S.\ Montanez and P.\ Resco,
{\it Semi-classical mechanics in phase space: the quantum target
of minimal strings},
[arxiv:hep-th/05066159].

                                            
\end{thebibliography}
\end{document}